\documentclass{article}[12pt]
\usepackage{indentfirst}
\usepackage{tabularx}
\usepackage[a4paper]{geometry}
\usepackage[utf8]{inputenc}
\usepackage[T1]{fontenc}
\usepackage{lmodern}
\usepackage{amsmath,amssymb,amsthm,centernot} \allowdisplaybreaks[1]
\usepackage{mathrsfs}
\usepackage{microtype}
\usepackage{latexsym}
\usepackage[colorlinks]{hyperref}
\usepackage{cleveref}
\usepackage{setspace}
\usepackage{graphicx}
\usepackage{bookmark}
\usepackage{booktabs}
\usepackage{longtable}
\usepackage{diagbox}
\usepackage{float}
\usepackage{color}   
\usepackage{bm}
\usepackage{multirow}
\usepackage{multicol}
\usepackage{arydshln}
\usepackage{makecell}
\usepackage{threeparttable}

\newtheorem{thm}{Theorem}[section]
\newtheorem{lm}{Lemma}[section]
\newtheorem{prop}{Proposition}[section]

\theoremstyle{definition}

\newcommand{\F}{\mathbb{F}}

\title{Some $3$-designs invariant under $2.P{\Sigma}L(2,49).$}
\author{ Minjia Shi\thanks{smjwcl.good@163.com},
	Ruowen Liu\thanks{liuruowen0116@163.com},
	Patrick Sol\'e\thanks{sole@enst.fr}
	\thanks{Minjia Shi and Ruowen Liu are with the Key Laboratory of Intelligent Computing Signal
		Processing, Ministry of Education, School of Mathematical Sciences, Anhui
		University, Hefei 230601, China; State Key Laboratory of integrated Service Networks, Xidian University, Xi'an,
		710071, China.  Patrick Sol\'e is with Aix Marseille Univ, CNRS, I2M, Marseilles, France. } }

\date{}
\begin{document}
	\maketitle
	\begin{abstract}
		We construct a ternary [49,25,7] code from the row span of a Jacobsthal matrix. It is equivalent to a Generalized Quadratic Residue (GQR) code  in the sense of van Lint and MacWilliams (1978). These codes are the abelian generalizations of the quadratic residue (QR) codes which are cyclic. The union of the [50,25,8] extension of the said code and its dual supports a 3-(50,14,1248) design. The automorphism group of the latter design  is a double cover of the permutation part of the automorphism group of the [50,25,8] code,
which is isomorphic to $P{\Sigma}L(2,49).$
Other weights in this code, other GQR codes, and other QR codes yield other 3-designs by the same process. A simple group action argument is provided to explain this behaviour of isodual codes.
	\end{abstract}
	\section{Introduction}\label{Introduction}
There has been a recent flurry of activity on designs supported by codes where different weight classes yield designs of different strengths \cite{AMMN,BS,I,MMN}. This is motivated by analogies with lattices \cite{M}. Thus the existence of these designs can be explained neither by the Assmus-Mattson theorem, nor by a transitivity argument, which give designs for all weights \cite{DT}. In \cite{BS}, a $3$-design satisfying this constraint was constructed from the codewords of weight $10$ of the binary extended quadratic residue code of length $42.$ Similar examples from ternary and quaternary quadratic residue codes were provided in \cite{I}. A symmetry explanation of the existence of the $3$-design on $42$ points was provided in \cite{AMMN}. In the same paper, a construction of $3$-designs from the codewords of given weight of a binary isodual code and its dual was given.
A general theorem on isodual binary codes with a permutation group having two orbits on triples was given in \cite[Th.1.1]{AMMN}.

In the present note, we consider a ternary generalized quadratic residue code of length 49 and its dual. Such a code is defined as the row span of a Jacobsthal matrix \cite{BSC}. It is also a principal ideal in the group ring $\F_3[\F_q]$, and is a Generalized Quadratic Residue (GQR) code in the sense of \cite{LM}. The extension code is invariant under the projective semi linear group $P{\Sigma}L(2,49)$ \cite{H}, and is isodual. We checked by machine computations \cite{M}, that the codewords of given weight $w$ with $w \in \{8,12,14,15,17,18\}$ in the extension code and its dual support a 3-design. This experimental fact would suggest that a ternary analogue of \cite[Th.1.1]{AMMN} exists. Indeed, we will provide an elementary analogue of that argument, which does not require Jacobi polynomials or Harmonic weight enumerators. This result shows that all weights of that code give $3$-designs.
In a second part of the paper, we apply the same construction technique of designs to the ternary extended GQR code of length $26,$ and to the ternary extended quadratic residue codes of lengths $14$ and $38.$

The material is arranged as follows. The next section collects the notions and notations needed in the rest of the paper. Section 3 introduces the codes and their properties. Section 4 contains the designs we found. Section 5 concludes the article.
\section{Definitions and notation}
\subsection{Ternary codes}

A ternary linear code is defined over the finite field $\F_3$, the Galois field with three elements. Formally, a ternary linear code $C$ of length $n$ and dimension $k$ is a $k$-dimensional subspace of the vector space $\F_3^n$ over $\F_3$. The minimum distance $d$ of a linear code $C$ is defined by:

\[
d(C) = \min\{d(x, y) \mid x, y \in C, x \neq y\},
\]

where $d(x, y)$ denotes the \textbf{Hamming distance} between two vectors $x$ and $y$, which is the number of coordinate positions in which $x$ and $y$ differ. The code $C$ can be described as an $[n, k, d]_3$ code, indicating its length $n$, dimension $k$, and minimum distance $d$. The \textbf{weight} $w(x)$ of a codeword is its distance to zero: $w(x) = d(x, 0)$. The weight distribution in Magma notation \cite{M}, is given by a list of ordered pairs $\langle i, A_i\rangle$ of the form $[\langle0,1\rangle,\cdots,\langle, A_i\rangle,\cdots]$ where only pairs with $A_i \ne 0$ are given. The \textbf{support} of a codeword $x$ of a code of length $n$ is the set of indices $i$ where $x_i \neq 0$:

\[S(x) = \{ i \in \{1,\dots, n\} \mid x_i \neq 0\}.\]

The {\bf permutation group} $Perm(C)$ of a ternary code $C$ is the group of all coordinate permutations  that leave the code wholly invariant.
The {\bf automorphism group} $Aut(C)$ of a ternary code $C$ is the group of all coordinate permutations and coordinate negations and their compositions, sometimes called {\bf monomial transforms} that leave the code wholly invariant.
The {\bf permutation part} $Per(C)$ of that group is the set of all permutations occurring in that group when negations are omitted. Note that the permutation group is a subgroup of the permutation part of the automorphism group. Two codes are \textbf{equivalent} if there is a monomial transform mapping one to the other.

{\lm \label{per} The group $Per(C)$ permutes the supports of the codewords of $C.$}

\begin{proof}
 Let $f \in Aut(C).$ Let $p: Aut(C) \to Per(C)$ be the map which associates to $f \in Aut(C)$ its permutation part.
 If $x \in C,$ then $x^f \in C.$ Now, since negations do not affect the support, we see that $S(x^f)=S(x)^{p(f)}.$ The result follows.
\end{proof}

The {\bf dual} $C^\perp$ of a code $C$ is understood with respect to the standard inner product $(.).$
$$C^\perp=\{x\in \F_3^n \mid \forall y \in C,\, (x.y)=0\}.$$
A ternary code is {\bf isodual} if it is equivalent to its dual, and {\bf self-dual} if it is equal to its dual.
A {\bf  cyclic} ternary code of length $n$ is an ideal in $ \frac{\F_3[x]}{(x^n-1)}$
of the form $(g(x)).$ When $n=p,$  an odd prime, choosing

$$g(x)=\prod_{i=\square \,\in \F_p} (x-\alpha^i),$$
 with $\alpha$ a root of unity of order $n$ in the algebraic closure of $\F_3$ yields
a {\bf quadratic residue code} (QR) \cite{MS}. Note that this polynomial has coefficients in $\F_3$ only if $3$ is a quadratic residue mod $p.$ This happens iff $p \equiv \pm 1 \pmod{12}.$

 A ternary code is {\bf abelian} if it is an ideal in the group ring $\F_3[G]$ where $G$ is an abelian group. This ring consists of the formal polynomials $\sum_{f \in G} a_f x^f$ in the undeterminate $x,$ with the componentwise addition
 \[ (\sum_{f \in G} a_f x^f)+(\sum_{g \in G} b_g x^g)=\sum_{g \in G}(a_g+ b_g) x^g\]

and the multiplication is defined by the convolution product
\[ (\sum_{f \in G} a_f x^f)(\sum_{g \in G} b_g x^g)=\sum_{h \in G}( \sum_{fg=h} a_fb_g) x^h.\]


\subsection{Designs}
A {\bf combinatorial design} of strength $t$ is a multiset $B$ of $K$-sets (called blocks) of a $v$-set of points $\Omega$ such that any $t$-tuple of $\Omega^t$ is contained in exactly $\lambda$ blocks. Its {\bf parameters} are denoted compactly as t-$(v,K,\lambda).$ If $B$ is a set then the design is said to be {\bf simple,} and we let $b=|B|.$ If, furthermore, $|B|={v \choose K}$, then the design is called {\bf trivial}.
The {\bf automorphism group} Aut$(D)$ of the design  $D$ is the set of permutation of points that leave $B$ wholly invariant.
Two designs $D_1$ and $D_2$ are {\bf isomorphic} if they share the point set $\Omega$ and there is a permutation of $\Omega$ that maps the blocks of $D_1$ to the blocks of $D_2.$
\subsection{Permutation groups}
A permutation group acting on a set $X$  is transitive if it has exactly one orbit on $X.$ It is  {\bf $t$-transitive} (resp. {\bf $t$-homogeneous}) if it is transitive in the induced action on ordered $t$-tuples
 $X^t$ (resp. $t$-subsets of $X$ that is ${X \choose t}$).
Recall that $GL(n,q)$ denote the general linear group, the group of n by n invertible matrices over $\F_q.$ The {\bf special linear group} $SL(n,q)$ is the group of matrices in $GL(n,q)$ of determinant unity. The {\bf projective linear group} $PSL(n,q)$ is the quotient of $SL(n,q)$ by scalar matrices. Recall that $B \propto A$ denotes the semi-direct product of the group $A$ extended  by the group $B.$ Thus the {\bf projective semi-linear group} $P{\Sigma}L(n,q) \simeq \text{Gal}(\F_q) \propto  PSL(n,q).$

\section{Isodual codes}

Let $C$ be an isodual code of length $n$ with $\sigma \in \text{Aut}(\F_3^n),$ such that $C^\sigma = C^\perp.$ Assume that $\text{Per}(C)$ has only two orbits on triples of $[n] = \{1, \dots, n\},$ that are exchanged by $p(\sigma),$ the permutation part of $\sigma.$ Thus
$${[n] \choose 3}^{\text{Per}(C)} = O_1 \sqcup O_2,$$
with $O_i^{p(\sigma)} = O_j$ for $i \neq j.$
Following Magma notation \cite{M}, we write
$$\text{Words}(C, w) = \{x \in C \mid w(x) = w\}.$$
The main result of this note is as follows.

{\thm \label{main} Keep the previous notation. Let $w$ be a nonzero weight of $C.$ Let
$$B = \text{Words}(C, w) \cup \text{Words}(C^\perp, w).$$
The elements of $S(B)$ are the blocks of a $3$-design on $[n].$
This design is invariant under the group generated by $\text{Per}(C)$ and $\sigma.$}

\begin{proof}
Let $T \in {[n] \choose 3}.$ Thus $T \in O_i$ for some $i \in [2].$
We need to show that the following number does not depend on $T$:
$$ \beta(T) = |\{ x \in B \mid T \subset S(x) \}|. $$

Consider two cases for an arbitrary $T' \in {[n] \choose 3}:$
\begin{itemize}
    \item If $T' \in O_i$, by action of $\text{Per}(C)$ we have $\beta(T) = \beta(T')$ since $\text{Per}(C)$ permutes $B$ by Lemma \ref{per}.
    \item If $T' \in O_j$ with $j \neq i$, then ${T'}^{p(\sigma)} \in B^\sigma = B.$
    Hence $\beta(T') = \beta({T'}^{p(\sigma)}) = \beta(T)$, since ${T'}^{p(\sigma)} \in O_i$.
\end{itemize}

The first statement follows. As for the second statement, it is clear that the sets $Words(C,w)$ and $Words(C^\perp,w)$ are exchanged by $\sigma$, and that both are invariant under $Aut(C)$. The second statement follows by considering permutation parts.
\end{proof}

\section { A family of ternary abelian Codes}

Let $q$ be an odd prime power. Let $\chi$ be the quadratic character of $\F_q^\times$ defined by
\[
\chi(x) = \begin{cases}
1, & \text{if } x = \square, \\
-1, & \text{if } x \neq \square,
\end{cases}
\]
and extended to $\F_q$ by the convention $\chi(0) = 0$. Consider the $q \times q$ matrix $W$ indexed by $\F_q$ with entries $W_{xy} = \chi(x - y)$. This matrix is instrumental in constructing Hadamard matrices of Paley type \cite{MS} and traditionally called the \textbf{Jacobsthal} matrix associated with $q$.

Then we consider $C(q)$, the row span of $W + I_q$ over $\F_3$, where $I_q$ denotes the identity matrix of order $q$. Denote by $ E(q)$ the extension of $C(q)$ by an overall parity-check.

{\prop The code \( C(q) \) in the case where \( q = p^2 \), with \( p \) being an odd prime \( \equiv 1 \pmod{3} \), is a specialization over \( \F_3 \) of the universal quadratic residue code described in \cite{C,BSC}.}

\begin{proof}
Consider the definitions in \cite[p.369, right column]{BSC}. Since \( q \equiv 1 \pmod{4} \), we have \(\epsilon = 1\) and \(\delta = p\), an integer \(\equiv 1 \pmod{3}\). The result follows.
\end{proof}

We write  $\F_3[\F_q]$ for the group ring $\F_3[(\F_q,+)].$ With this notation,
the following Proposition is then immediate from the definition of $E(q).$ Its proof is omitted.

{\prop
Let $U$ (resp. $V$) denote the set of nonzero squares (resp. nonsquares) in $\F_q.$
The code $C(q)$ is a principal ideal in the ring $\F_3[\F_q]$ with generator $ x^0+\sum_{f \in U} x^f-\sum_{f \in V} x^f .$}
\\ \ \

The following fact can be verified in Magma \cite{M}.

{\prop The code $C(49)$ is equivalent to the GQR code of idempotent generator $x^0+\sum_{f \in U} x^f. $}
\\ \ \

The next result follows by \cite{H}.

{\prop \label{fact3} The permutation part of the automorphism group of the code $E(49)$ is isomorphic to $P{\Sigma}L(2,49).$}

\begin{proof} Follows by \cite[Th. 5.3 (v) a)]{H}.
\end{proof}
\section{The extended GQR code of length $50$}

 The following facts can be verified easily in Magma \cite{M}. We give a computer free proof for some of them.
{\thm
 Using Jacobsthal matrix to construct $E(49)$, we have:
\begin{enumerate}
\item[(1)] $E(49)$ is a $[50,25,8]$ isodual code.
\item[(2)] The permutation part of the automorphism group of $E(49)$ is isomorphic to $P{\Sigma}L(2,49),$ of order $117600.$
\item[(3)] Let $B$ be the union of codewords of $E(49)$ and its dual. Then the vectors of $B$ of weights $8,12,14,15,17,18$ hold $3$-designs with the parameters in Table 1.
\item[(4)] These $3$-designs are invariant under a double cover of $P{\Sigma}L(2,49),$ of order $235 200.$ This group is not $3$-homogeneous.
\item[(5)] A partial weight distribution of $E(49)$ is $$[\langle0,1\rangle,\langle8,350\rangle,\langle12,14700\rangle,\langle14,67200\rangle,\langle15,67200\rangle,\langle16,470400\rangle,\langle17,3247230\rangle,$$ $$\langle18,10923472\rangle,\langle19,346265236\rangle,\cdots].$$

\end{enumerate}}

\begin{proof}\noindent
\begin{enumerate}
\item[(1)] The minimum distance ($=7$) of $C(49)$ follows by invoking \cite[Theorem 2,(iii)]{LM}. Hence the minimum distance of $E(49)$ is at most $8.$ Isoduality follows by \cite[Lemma 4]{LM}.
\item[(2)] The permutation part of the automorphism group of $E(49)$ is  $P{\Sigma}L(2,49),$ by Proposition \ref{fact3}.
\item[(3)] This follows by Theorem \ref{main}, upon checking that $P{\Sigma}L(2,49),$
has two orbits on triples.
\item[(4)] Invariance follows by the second statement of Theorem \ref{main}. The fact that
this group is not $3$-homogeneous follows by \cite[Th. 1, (ii)]{K} and $49 \equiv 1 \pmod{4}.$
\item[(5)] Computer-assisted proof.
\end{enumerate}
\end{proof}

{\bf Remarks:}
\begin{itemize}
\item For lack of computer resources we could not check the weights > 18.
\item The permutation group of $E(49)$ is of order $2352=\frac{117600}{50}.$
\item There is a design with parameters 3-(50,8,1) in La Jolla covering repository \cite{LJ}, constructed using widely different methods. It is isomorphic to the design in Table 1. It would be interesting to determine if a design with these parameters is unique.

\item None of these designs belong to one of the infinite families of 3-designs of \cite[ \S 4.37, p.82]{CD}.
\end{itemize}

\begin{table}[h]
\setlength\tabcolsep{2pt}
\centering
\begin{tabular}{|c|c|c|c|c|c|c|}

\multicolumn{6}{c}{\textbf{Table 1 : 3-designs from E(49)}} \\
\hline
 w& 8 & 12 & 14 & 15 & 17 &18 \\
\hline
\text{parameters} &3-(50, 8, 1) &3-(50, 12, 165) &3-(50, 14, 1248) &3-(50, 15, 1560)& 3-(50, 17, 57800)& 3-(50, 18, 248370) \\
\hline
b &350  & 14700 &67200 & 67200& 1666000& 5965750  \\
\hline
\end{tabular}
\end{table}
\section{Other codes}
\subsection{The extended GQR code of length $26$}

The code $E(25)$ has for permutation part of its automorphism group $P{\Sigma}L(2,25)$ by Proposition \ref{fact3}. This group is neither $3$-transitive nor $3$-homogeneous by \cite{K}. But it has two orbits on triples.

The weight distribution of $E(25)$ is $$[ \langle0, 1\rangle, \langle6, 130\rangle, \langle8, 650\rangle, \langle10, 3510\rangle, \langle11, 9360\rangle, \langle12, 24700\rangle, \langle13, 55200\rangle, \langle14, 102700\rangle,$$
 $$\langle15, 154960\rangle, \langle16, 219180\rangle, \langle17, 250900\rangle, \langle18, 263900\rangle, \langle19,210600\rangle, \langle20, 156390\rangle, $$
 $$\langle21, 84500\rangle, \langle22, 42900\rangle, \langle23, 10400\rangle, \langle24, 3250\rangle, \langle25,780\rangle, \langle26, 312\rangle ].$$

All these weights yield $3$-designs as can be seen from Tables 2, 3, and 4. The weights $\geq 23$ yield trivial designs and are therefore omitted.

\begin{table}[h]
	\setlength\tabcolsep{2pt}
	\centering
	\begin{tabular}{|c|c|c|c|c|c|}
	\multicolumn{6}{c}{\textbf{Table 2 : 3-designs from E(25)}} \\
		\hline
		w & 6 & 8 & 10 & 11 & 12 \\
		\hline
		\text{parameters} & 3-(26, 6, 1) & 3-(26, 8, 14) & 3-(26, 10,162 ) & 3-(26, 11, 594) & 3-(26, 12, 1980 )  \\
		\hline
		b & 130 &  650& 3510 & 9360 &  23400 \\
		\hline
	\end{tabular}
\end{table}

\begin{table}[h]
	\setlength\tabcolsep{2pt}
	\centering
	\begin{tabular}{|c|c|c|c|c|c|}
	\multicolumn{6}{c}{\textbf{Table 3 : 3-designs from E(25)}} \\
		\hline
		w & 13 & 14 & 15 & 16 & 17 \\
		\hline
		\text{parameters} & 3-(26, 13, 6072) & 3-(26, 14, 13923) & 3-(26, 15, 27118) & 3-(26, 16, 43246) & 3-(26, 17, 60180)  \\
		\hline
		b & 55200 & 99450 & 154960 & 200785 & 230100  \\
		\hline
	\end{tabular}
\end{table}

\begin{table}[h]
	\setlength\tabcolsep{2pt}
	\centering
	\begin{tabular}{|c|c|c|c|c|}
		
		\multicolumn{5}{c}{\textbf{Table 4 : 3-designs from E(25)}} \\
		\hline
		w & 18 & 19 & 20 & 22 \\
		\hline
		\text{parameters} &  3-(26, 18, 68544) & 3-(26, 19, 51357) & 3-(26, 20, 49077) & 3-(26, 22, 4235) \\
		\hline
		b & 218400 & 137800 & 111930 & 7150 \\
		\hline
	\end{tabular}
\end{table}

\subsection{The extended QR codes of length $14$ and $38$}
QR codes over $\F_3$ are studied in some detail in \cite[Chap. 16, \S8]{MS}. A parameter table is shown in Fig. 16.2(b) \cite[p. 483]{MS}. They are known to be isodual when $p \equiv 1 \pmod{4}$ \cite[p. 482]{MS}. This condition is satisfied for our construction of $3$-designs. We specifically consider the primes $p = 13$ and $p = 37$, as $p = 61$ is too large for computational exploration.

Denote the extended QR code of length $p+1$ by $ XQR(p)$. The two codes $XQR(13)$ and $XQR(37)$ have a permutation part of their automorphism group as $PSL(2, 13)$ and $PSL(2, 37)$, respectively. Note that $PSL(2, q)$ is always $2$-transitive but $3$-homogeneous only when $q \equiv 3 \pmod{4}$ \cite{DT}. Both $PSL(2, 13)$ and $PSL(2, 37)$ have two orbits on triples, as argued in \cite{DT}.

\subsubsection{The extended QR code of length 14}
 The weight distribution of $XQR(13)$ is
$$[ \langle0, 1\rangle, \langle6, 182\rangle, \langle7, 156\rangle, \langle8, 364\rangle, \langle9, 364\rangle, \langle10, 546\rangle, \langle11, 364\rangle, \langle12,182\rangle, \langle14, 28\rangle ].$$

All these weights < 11 yield $3$-designs as can be seen from Table 4. The weights $11,12,14$ yield no design.

\begin{table}[h]
	\setlength\tabcolsep{2pt}
	\centering
	\begin{tabular}{|c|c|c|c|c|c|}
		
		\multicolumn{6}{c}{\textbf{Table 4 : 3-designs from XQR(13)}} \\
		\hline
		w & 6 & 7 & 8 & 9 & 10 \\
		\hline
		\text{parameters} &  3-(14, 6, 10)  & 3-(14, 7, 15) & 3-(14, 8, 42) & 3-(14, 9, 84) & 3-(14, 10, 90) \\
		\hline
		b & 182 & 156 & 273 & 364 & 273 \\
		\hline
	\end{tabular}
\end{table}

The $3$-design of parameters 3-(14, 10, 90) is to be expected from Theorem 1.1 of \cite{I}, which shows that the codewords of weight 10 of $XR(13)$ form a $3$-design.
\subsubsection{The extended QR code of length 38}
 The weight distribution of $XQR(37)$ is $$[ \langle0, 1\rangle, \langle11, 2812\rangle, \langle12, 12654\rangle, \langle13, 25308\rangle, \langle14, 156066\rangle, \langle15, 421800\rangle, \langle16,1290708\rangle,$$
$$\langle17, 3180372\rangle, \langle18, 7565686\rangle, \langle19, 15918732\rangle, \langle20, 30569252\rangle, \langle21,51662064\rangle,$$
$$\langle22, 80441478\rangle, \langle23, 111186480\rangle, \langle24, 140088216\rangle, \langle25, 156074436\rangle,\langle26, 156719790\rangle, $$
$$\langle27, 138586608\rangle,\langle28, 109241982\rangle, \langle29, 75161948\rangle, \langle30,45419424\rangle, \langle31, 23283360\rangle, $$
$$\langle32, 10186470\rangle, \langle33, 3619044\rangle, \langle34, 1164168\rangle, \langle35,236208\rangle, \langle36, 44992\rangle, \langle38, 1408\rangle ].$$

All these weights < 16 yield $3$-designs as can be seen from Table 5. Due to limited computer resources, we were unable to verify weights $16$ and more.

\begin{table}[h]
	\setlength\tabcolsep{2pt}
	\centering
	\begin{tabular}{|c|c|c|c|c|c|}
		
		\multicolumn{6}{c}{\textbf{Table 5 : 3-designs from XQR(37)} } \\
		\hline
		w & 11 & 12 & 13 & 14 & 15 \\
		\hline
		\text{parameters} &  3-(38, 11, 55)  & 3-(38, 12, 330) & 3-(38, 13, 858) & 3-(38, 14, 6734) & 3-(38, 15, 22750) \\
		\hline
		b & 2812 & 12654 & 25308 & 156066 &421800 \\
		\hline
	\end{tabular}
\end{table}

\section{Conclusion} In this note, we present some $3$-designs supported by  certain codes and their duals. Specifically, we have studied the extended ternary GQR code of length 50 and its dual, along with the extended GQR code of length 26, and the extended QR codes of lengths 14 and 38. As there is no Assmus-Mattson theorem applicable to designs supported jointly by a code and its dual, the designs constructed here cannot be explained solely by weight properties. Additionally, the permutation group of these codes is neither 3-transitive nor 3-homogeneous. Thus, the standard transitivity argument does not account for their existence.

The explanation provided in \cite{AMMN} for binary isodual codes has been generalized and simplified to ternary codes. As its application to GQR codes is based on the number of orbits of $P\Sigma L(2,q)$ on triples, it would be interesting to have a  proof for general $q$ that the number of these orbits is two, and that they are exchanged by the permutation part of the transform that maps extended GQR codes on their duals.

\end{document}